\documentclass[aps,twocolumn,prl,showpacs,color,psfig,epsf]{revtex4}
\usepackage{amsmath}
\usepackage{color}
\usepackage{amsfonts}
\usepackage{epsf}
\usepackage{graphicx}
\begin{document}

\title{Self-assembly and nonlinear dynamics of dimeric colloidal rotors in cholesterics}

\author{J. S. Lintuvuori$^1$, K. Stratford$^2$, M. E. Cates$^1$, D. Marenduzzo$^1$}
\affiliation{$^1$SUPA, School of Physics and Astronomy, University of Edinburgh, Mayfield Road, Edinburgh, EH9 3JZ, UK;\\
$^2$EPCC, School of Physics and Astronomy, University of Edinburgh, Mayfield Road, Edinburgh, EH9 3JZ, UK.}

\begin{abstract}
We study by simulation the physics of two colloidal particles in a cholesteric liquid crystal with tangential order parameter alignment at the particle surface. The effective force between the pair is attractive at short range and favours
assembly of colloid dimers at specific orientations relative to the local director field. When pulled through the fluid by a constant force along the helical axis, we find that such a dimer rotates, either continuously or stepwise with phase-slip events.
These cases are separated by a sharp dynamical transition and lead respectively to a constant or an ever-increasing phase lag between the dimer orientation and the local nematic director.
\end{abstract}

\maketitle

Self-assembly is one of the key aims of present-day nanotechnology. The idea underpinning this concept is that, through a careful control of the interactions in a suspension of particles, it is possible to drive the spontaneous formation of an ordered target structure or pattern, starting from a disordered initial condition. The process is spontaneous as it often entails the minimisation, whether global or local, of the free energy of the system.
An outstanding specific challenge in modern self-assembly is to build a structure with a target {\it dynamic} feature, such as a synthetic microscopic rotor, walker or swimmer~\cite{colloidal_walkers}. 

Colloidal dispersions in liquid crystals offer a useful system for the development of self-assembly strategies at the microscale.
Even in nematics, the simplest liquid crystalline phase, elastic distortions and topological defects mediate a variety of interactions resulting in the formation of wires, colloidal crystals, cellular solids, and clusters entangled by disclinations~\cite{nematic_self_assembly,Tanaka,zumer_colloid}.
This variety of self-assembled structures is possible because one can tune the liquid crystal mediated interactions by changing, e.g., the strength and nature of the liquid crystal ordering at the colloidal surface.
These alter the local symmetry of the director field near the particle and hence qualitatively change the effective interparticle forces.  

Here we show that a powerful way to extend the potential for self-assembly of colloids in liquid crystals is to consider cholesterics, or chiral nematics.
In cholesterics the direction of the molecular order in the ground state spontaneously twists.
The spatial modulation of the twist is in the micron range, therefore of the order of typical colloidal sizes. We have previously shown~\cite{juho} that by varying the ratio between these two fundamental length scales, it is possible to change continuously the topology of the defects, or disclination lines, surrounding a single particle---morphological changes with no direct counterpart in nematics. 
Very recently, Mackay and Denniston~\cite{colin} took a step further and studied the interparticle elastic force felt in a cholesteric by two colloidal spheres with tangential anchoring of the director field at their surface. As in the nematic case~\cite{lavrentovich_planar}, there is a complex interplay between repulsive and attractive directions which leads to the formation of a dimer or longer chains.
Here we focus on the simplest case of a dimer, but progress
beyond the purely static investigation of~\cite{colin} to show that such a dimer
exhibits unexpected and intriguing dynamical properties.
When subjected to an external force (for instance gravity)
along the cholesteric helix,
the dimer rotates about this axis in a screw-like fashion.
Depending on the magnitude of the force, the dimer either rotates
continuously or exhibits phase slippage, alternating periods of smooth
rotation with static spells in which it translates without rotation.
This dynamical transition shows similar near-critical behaviour to the depinning of driven vortices and of charged density waves in superconductors, both of which may be studied with the Frenkel-Kontorova model for transport in a periodic potential~\cite{frenkel-kontorova}. Another analogue is provided by the synchronization of coupled oscillators described by the Kuramoto model~\cite{kuramoto}.
Within our liquid crystal context, the 
phase slippage regime requires a specific free energy landscape which we
discuss.
This provides potential for the design of self-assembled systems with tunable
dynamic properties.

The system we study consists of two spherical colloidal particles of radius
$R$ moving in a cholesteric liquid crystal.
To describe the thermodynamics of the chiral host, we employ a Landau--de Gennes free energy $\cal{F}$, whose density ${f}$ may be expressed in terms of a traceless and symmetric tensor order parameter $\mathbf{Q}$~\cite{beris} and is
detailed in~\cite{supmat}.
\if{
\begin{align}
{f} & = \tfrac{A_0}{2} \bigl( 1 - \tfrac{\gamma}{3} \bigr) Q_{\alpha \beta}^2 
           - \tfrac{A_0 \gamma}{3} Q_{\alpha \beta}Q_{\beta \gamma}Q_{\gamma \alpha}
           + \tfrac {A_0 \gamma}{4} (Q_{\alpha \beta}^2)^2 \notag \\ 
	 & \quad + \tfrac{K}{2}\bigl( \nabla_{\beta}Q_{\alpha \beta}\bigr)^2
	   + \tfrac{K}{2} 
           \bigl( \epsilon_{\alpha \gamma \delta} \nabla_{\gamma} Q_{\delta \beta} 
           + 2q_0 Q_{\alpha \beta} \bigr)^2. 
\label{eq:FreeEnergy}
\end{align}
}\fi
\if{
Here, $A_0$ sets the energy scale, $K$ is an elastic constant, $q_0=2\pi/p$ where $p$ is the cholesteric pitch, and $\gamma$ is a temperature-like control parameter governing proximity to the isotropic-cholesteric transition.
Greek indices denote Cartesian components and summation over repeated indices is implied; $\epsilon_{\alpha\gamma\delta}$ is the permutation tensor.
}\fi
Tangential anchoring is modelled by a surface free energy, $f_s= {\textstyle \frac{1}{2}} W (Q_{\alpha\beta}-Q^0_{\alpha\beta})^2$, where $W$ is the strength of anchoring and $Q^0_{\alpha\beta}$ is the preferred order parameter in the tangent plane to the local spherical surface~\cite{galatola}.

We employ a 3D hybrid lattice Boltzmann (LB) algorithm~\cite{LBLC} to solve the Beris-Edwards equations for $\mathbf{Q}$~\cite{beris}
\begin{equation}
D_t \mathbf{Q} 
= \Gamma  \Bigl( \tfrac{-\delta {\cal F}}{\delta \mathbf{Q}} + \tfrac{1}{3}\, 
\text{tr} \Bigl( \tfrac{\delta {\cal F}}{\delta \mathbf{Q}} \Bigr) \mathbf{I} \Bigr)  .
\label{eqQevol}
\end{equation} 
Here, $\Gamma$ is a collective rotational diffusion constant and $D_t$ is the material derivative for rod-like molecules~\cite{beris}. The term in brackets is known as the molecular field, which in the absence of flow drives the system towards a free energy minimum. The boundary conditions for the order parameter on the colloidal surfaces are given by~\cite{zumer_colloid}:
\begin{equation}
\nu_{\gamma}
\frac{\partial f}{\partial \partial_{\gamma} Q_{\alpha\beta}}
+\frac{\partial f_s}{\partial Q_{\alpha\beta}}=0
\end{equation}
where $\nu_{\gamma}$ is the local normal to the colloid surface.

The velocity field obeys the continuity and Navier-Stokes equation, with a stress tensor generalised to describe liquid crystal hydrodynamics~\cite{beris}. Within our hybrid scheme, we solve the Navier-Stokes equation via LB, and Eq.~\ref{eqQevol} via finite difference~\cite{juho}.
Colloids are represented by the standard method of bounce-back on links, which leads to a no-slip boundary condition for the velocity field (see~\cite{juho,ladd} for details). Order parameter variations create an additional elastic force acting on the particle
which is computed by integrating the stress tensor over the particle surface~\cite{juho,zumer_colloid}.

The dynamics is primarily controlled by the Ericksen number, ${\mathrm{Er}}={\gamma_1 vR}/{K}$, where the rotational viscosity $\gamma_1={2q^2}/{\Gamma}$, $v$ is a velocity characteristic of the flow, and $q$ is the  degree of ordering in the system. In the uniaxial case with director $\hat{\bf n}$, $Q_{\alpha\beta}=q(\hat{n}_\alpha \hat{n}_\beta-\delta_{\alpha\beta}/3)$. 

In what follows, we quote our results in simulation units~\cite{supmat,LBLC}. To convert them into physical ones, we can specify an elastic constant of 28.6 pN, and a rotational viscosity of 1 poise. (These values hold for typical materials, and a colloidal diameter of 1 $\mu$m.) In this way, the simulation units for force, time and velocity can be mapped onto 440 pN, 1 $\mu$s, and 0.07$\mu$m/s respectively. 


We first consider the interparticle elastic potential. 
Two particles are placed centre-centre separation vector $\mathbf{d}$ apart. 
Both particles are
held fixed for the duration of a simulation in which the free energy is
minimised. By repeating simulations for different $\mathbf{d}$, we map
out the effective two-body potential as a function of the reduced separation
$(d - 2R)/p$ in the three coordinate directions, and as a function of angle
in the $x-y$ plane (Fig.~1a).
The potential is markedly anisotropic. While the potential in $z$ shows strong repulsion at
large separations, there is an attraction in the $x-y$ plane. The most favourable
configuration at small separations is along the director field ($x$-direction in Fig. 1a). Here we estimate a maximum attractive force of 20 pN.
Before this deep minimum is reached the dimer needs to overcome a repulsion (peak
at $(d - 2R)/p \approx 0.2$), which is largest when the disclinations at the opposing particle surfaces join up.
In this bound state, the colloids share two disclination lines which act as a glue
between them (right Fig. 1b). For separation vectors perpendicular to $\hat{\mathbf n}$ we find a stable
minimum and no repulsive barrier ($y$-direction in Fig. 1a).
These results are far from the nematic limit studied experimentally
in~\cite{lavrentovich_planar} and theoretically in~\cite{mozaffari},
as well as from results obtained in a twisted nematic cell~\cite{twisted_nem}.
Most notably, the in-plane potential perpendicular to $\hat{\mathbf n}$
(here $y$) was always repulsive in the nematic~\cite{mozaffari}.
The preferred configuration (here,
along $\hat{\mathbf n}$) is
at an angle of about 30$^{\circ}$ in the nematic~\cite{lavrentovich_planar,mozaffari}. 

\begin{figure}
\includegraphics[width=\columnwidth]{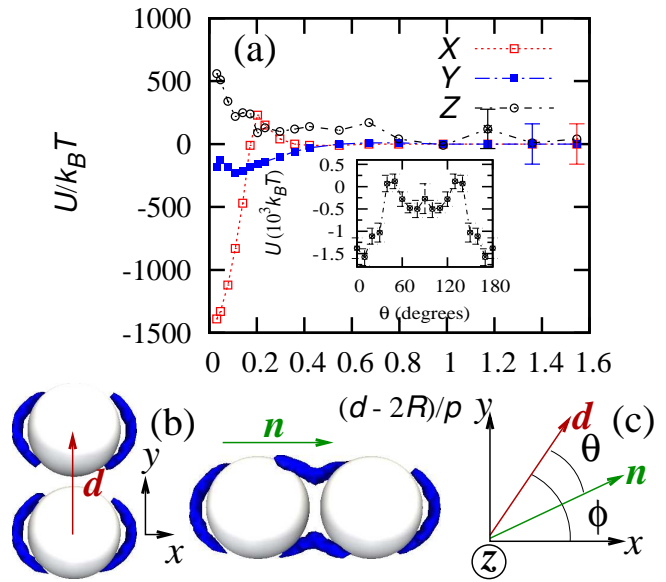}
\caption{[Color online.] (a) Interparticle elastic potential as a function of
reduced separation $(d - 2R)/p$,  along the helical axis ($z$) and in the
$x-$ and $y-$ directions. The inset shows the angular
dependence in the $x-y$ plane at $(d - 2R)/p \approx 0.05$.
Units are in $k_BT \approx 4$ pN nm.
(b) Snapshots of the dimer showing the disclination lines at
$(d - 2R)/p \approx 0.05$ when $\mathbf{d}$ is along $y$ and along $x$.
The Cartesian axes, the separation $\mathbf{d}$,
 and the far-field nematic director,
$\hat{\mathbf n}$ are shown. (c)
Definition of the angles $\theta$ and $\phi$ used in the text.
For error analysis see Supplementary Material~\cite{supmat}.}
\label{potential}
\end{figure}


These results confirm and extend those of~\cite{colin} on the
energetics of dimer formation in cholesterics. Our key focus in the present
work is dynamics.
We place two particles initially near the weaker minimum of the potential
in the direction perpendicular to $\hat{\mathbf n}$ (left Fig. 1b). 
This is the first relatively deep local minimum two particles approaching from far away would encounter. It is therefore a natural self-assembled configuration for the dimer.
We then pull each along the helical axis with force $f$.

At all force levels studied here,  the moving dimer
rotates about its centre of mass while $\mathbf{d}$ remains
perpendicular to the helical axis (inset Fig. 2e). 
The behaviour at low force ($f\le 0.025$) is illustrated in Fig.~2(a--d).
We quantify the rotation by measuring the angle, $\phi$, between $\mathbf{d}$ and the
$x$-axis and the angle, $\theta$, between $\mathbf{d}$
and $\hat{\mathbf{n}}(z)$  (Fig. 1c), as a function of
time. After an initial transient, a smooth rotation is observed (Fig.~2e; open symbols), 
but with separation ${\mathbf d}$ that lags behind
$\hat{\mathbf n}(z)$ by a constant phase angle $\theta(t)$ (Fig.~2f, open symbols)~\cite{note}. We attribute this
constant lag to a balance between the viscous drag opposing the
rotation of the pair in the $x-y$ plane, and the force arising from
the angular variation in the rotating interparticle potential. 

\begin{figure}
\includegraphics[width=\columnwidth]{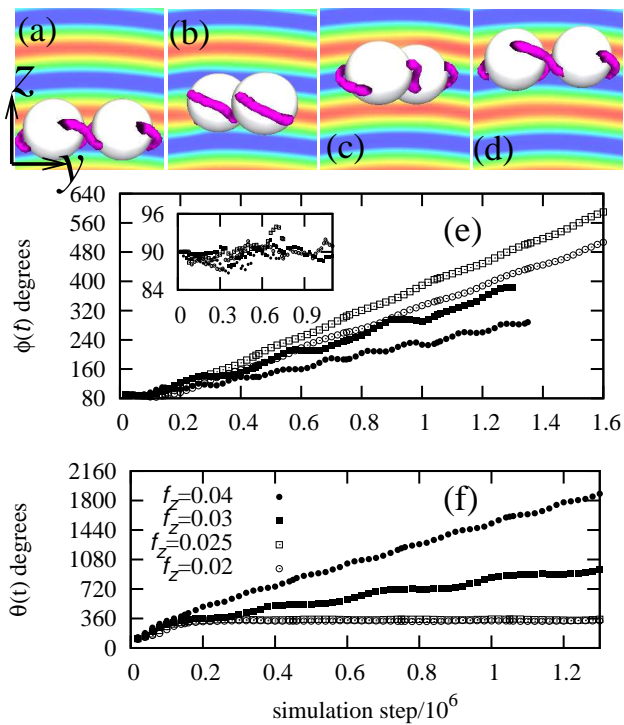}
\caption{[Color online.] (a-d) Snapshots of a steadily rotating dimer over a translation of half a pitch (a rotation of 180$^{\circ}$). The director field is colour coded
according to the local director (red along $x$, blue along $y$).
The time evolution of (e) $\phi$ and (f) $\theta$ (defined in Fig. 1C), for the rotor (open symbols) and phase slippage (closed symbols) motion. The inset in (e) also shows the angle between ${\mathbf d}$ and the helical-axis, which remains
close to 90$^\circ$.}
\label{ratchet}
\end{figure}

At higher force ($f\ge 0.0275$), the behaviour is
manifestly different. The angle between
$\mathbf{d}$ and $x$, shown in Fig.~2e (closed symbols),
now increases with time in a series of steps. These steps correspond to intervals
with and without significant rotation of the dimer as it moves along the
helix (see~\cite{supmat} for a movie  and further discussion of the dynamics). 
Correspondingly, the phase lag $\theta(t)$ fails to
reach a steady value (Fig.~2f; closed symbols), but increases indefinitely. 
We refer to this as
``phase slippage''. There is a clear transition between the smooth
``rotor'' regime and the phase slippage regime.

Fig.~3a quantifies the
dependence of the average rotational velocity of the dimer, $\Omega$,
on the applied force for simulations at a range of force values. It
can be seen that $\Omega$ is proportional to $f$ in the low force rotor
regime, while it decreases in the high force phase slippage regime.
Fig.~3b shows the dependence of the average speed of the dimer along
the helical axis with force, again showing a transition. However, it is
difficult to relate these transitions to the equilibrium potential which is strictly valid only at Er $=0$. 
Although the largest Eriksen number
remains low (Er $\approx  0.026$), 
the potential is likely to
be affected by dynamical effects including the local bending of the cholesteric layers (visible in
Fig.~2a-d~\cite{supmat}). The local bending of the layers for $\mathrm{Er}\ll1$, was also observed for a single colloid moving along the helical axis~\cite{juho}.

To understand the dynamical transition better, we write down a set of phenomenological equations for the evolution of the dimer position, $z(t)$, and for $\phi(t)$, which gives the direction of ${\mathbf d}$ in the lab frame (Fig. 1c).
We assume that the basic features of the cholesteric ordering may be captured by an effective angular potential, also periodic in $z$: $V(\phi,z)$. Our
equations read as follows,
\begin{eqnarray}\label{simple-theory}
\frac{d\phi}{dt} &  = & -\frac{1}{\gamma_{\phi}} \frac{\partial V(\phi,z)}{\partial \phi} \\ \nonumber
\frac{dz}{dt} &  = & - \frac{1}{\gamma_{z}} \frac{\partial V(\phi,z)}{\partial z} \\ \nonumber
V(\phi,z) & = & -A \cos\left[B\left(\phi-q_0z\right)\right] +fz.
\end{eqnarray}
Here, $A$ (units of energy) and $B$ (dimensionless) are positive constants, $f$ is the external forcing, while $\gamma_{\phi}$ and $\gamma_{z}$ are relaxational constants related to the rotational and translational friction of the dimer, and whose exact values we will not need.
Eqs.~\ref{simple-theory} may be solved by an ansatz suggested by the behaviour in Fig.~2f. We write $\phi(t) = \Omega t+\theta(f)$, where the director-dimer
angle follows the most favourable orientation apart from a phase lag,
$\theta(f)$.
This ansatz is a solution provided that $f$ is smaller than a critical threshold $f_c$, i.e. in the rotor phase. 
By estimating the average terminal velocity of the dimer as $f/\gamma_{\phi}$, we find that in the rotor phase the angular velocity is $\Omega\sim 2\pi f/(\gamma_z p)$ (this is true within statistical error using data in Fig. 3), and that the critical force is $f_c=A\gamma_z p/(2\pi\gamma_{\phi})$. Above this threshold, $\theta$ increases with time as in Fig. 2f (solid symbols). The asymptotic velocity $\Omega_{\rm lag}(= d\theta(t)/dt)\sim \sqrt{f^2-f_c^2}$~\cite{supmat} for $f\to f_c$ provides a useful ``order parameter'' to characterise the rotor-slippage transition. In contrast, the angular velocity $\Omega$ decreases as $1/f$ for large $f$. 

Importantly, Eqs.~\ref{simple-theory} capture both the near-critical behaviour of $\Omega_{\rm lag}$, and the large $f$ behaviour of $\Omega$ shown by our full LB simulations (see Fig. 3a for an $\Omega \sim 1/f$ fit and Fig. 3c for a $\Omega_{\rm lag}\sim  \sqrt{f^2-f_c^2}$ fit).
The near-critical behaviour of $\Omega_{\rm lag}$ in our transition is similar to that of the velocity of a chain of driven particles in a periodic potential described by the Frenkel-Kontorova model~\cite{frenkel-kontorova}, suggesting that our dimer provides a liquid crystal representative of a wider class of models~\cite{supmat}. Another analogue is with the synchronisation of two driven oscillators described by the Kuramoto model~\cite{kuramoto,supmat}, where synchronised and unsynchronised states correspond to the rotor and slippage regimes respectively.
At the same time, we note that the physics of our cholesteric dimers is richer than that in Eqs.~\ref{simple-theory}. Whereas the symmetry of the problem suggests that there should always be a very low force regime in which the dimer behaves as a rotor, the existence of the phase slippage regime at higher forces depends on the form of the effective pair potential.
The required conditions hold for tangential anchoring but, according to our  preliminary studies is violated in the normal anchoring case, for which we have not observed a phase slip regime.
 
Our study may be viewed as a generalisation of a classical problem: the sedimentation of two spheres in a viscous fluid. 
Intriguingly, Fig.~3b shows that the velocity--force curves are not linear, even in the Stokes limit of effectively zero Reynolds number. Rather, they appear to have different slopes (i.e. effective viscosities) in the rotor and phase slippage regimes. This is different to what occurs for a single sedimenting particle in a cholesteric, which leads to a linear velocity--force relation in the force range simulated here~\cite{juho}. This biphasic force-velocity curve is due to the dynamic transition we discussed, and has no counterpart in classical sedimentation, in either Newtonian or Maxwell fluids. In the Newtonian case, sinking side by side speeds up the particles, by up to a factor of 2~\cite{2sphere_sedimentation}, which is not true in our rotor phase (where we find that a dimer sediments {\em slower} than a single particle). In a Maxwell fluid, the repulsive or attractive interaction between two spheres sedimenting side-by-side is controlled by normal stresses~\cite{two-point-microrheology}. In our case, we find a novel velocity--dependent torque and a dramatic dependence on the nature of the anchoring.

\begin{figure}
\includegraphics[width=0.9\columnwidth]{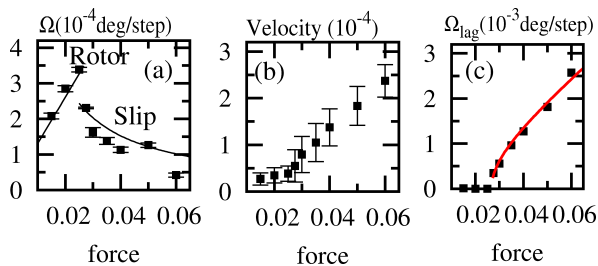}
\caption{Average rotational velocity ($\Omega$) (a) and sedimentation velocity (b) of the colloidal dimer as a function of the forcing. The non-linear fit in (a) is $\Omega=c/f$, with $c>0$ a constant. (c) shows $\Omega_{\rm lag}$ as a function of $f$ together with a fit to $d\sqrt{f^2-f_c^2}$, with $d>0$ a constant and $f_c\approx 0.0267\pm 0.0005$.}
\label{different-pulling-dynamics}
\end{figure}

In conclusion, we have studied the equilibrium and dynamic properties of two colloidal spheres in a cholesteric liquid crystal. We have seen that chirality leads to a major change in the effective potential felt by the pair, with respect to the nematic limit.  The elastic forces we find lead to the stabilisation of a dimer at an angle of either 0$^{\circ}$ or 90$^{\circ}$ degrees, as opposed to the 30$^{\circ}$ found in nematics. These results, alongside those of Ref.~\cite{colin}, suggest that it would be instructive to repeat the experiments performed in Ref.~\cite{lavrentovich_planar} with a cholesteric liquid crystal. Based on our results, one may also speculate that variations in particle size (or cholesteric pitch) can affect the local free energy landscape of colloidal suspensions in liquid crystals, and potentially drive the self-assembly of different structures.

Our main result is that, when subjected to an external force, the dimer rotates, either smoothly as a corkscrew or intermittently, with phase slippage. This transition occurs as the forcing exceeds a critical threshold: its value may be estimated via a simple theory considering the interplay between a spatially periodic angular potential and an external driving. The existence of the phase slippage regime is however highly non-trivial and relies on a delicate balance in the equilibrium and dynamic properties of our dimers: for example we have not observed it for dimers with normal anchoring. 
An interesting possibility for future research would be to study the effects of external electric field applied to our colloidal dimer, as done with platelets~\cite{platelets_electric} and nematic colloids~\cite{lavrentovich_backflow}, where unusual dynamics was triggered by the field. 

This work was funded by EPSRC Grants EP/E030173 and EP/E045316. MEC is funded by the Royal Society.

\onecolumngrid
\newpage

\makeatletter 
\def\tagform@#1{\maketag@@@{(S\ignorespaces#1\unskip\@@italiccorr)}}
\makeatother

\makeatletter \renewcommand{\fnum@figure}
{\figurename~S\thefigure}
\makeatother

\setcounter{equation}{0}
\setcounter{figure}{0}

\begin{center}
  {\Large \bf  Supplementary information for: Self-assembly and nonlinear dynamics of dimeric colloidal rotors in cholesterics}
\end{center}
\medskip

\section{Computational model and details}
To describe the thermodynamics of the chiral liquid crystal host, we employ a Landau--de Gennes free energy $\cal{F}$, whose density ${f}$ may be expressed in terms of a traceless and symmetric tensor order parameter $\mathbf{Q}$~\cite{beris1} as
\begin{align}
{f} & = \tfrac{A_0}{2} \bigl( 1 - \tfrac{\gamma}{3} \bigr) Q_{\alpha \beta}^2 
           - \tfrac{A_0 \gamma}{3} Q_{\alpha \beta}Q_{\beta \gamma}Q_{\gamma \alpha}
           + \tfrac {A_0 \gamma}{4} (Q_{\alpha \beta}^2)^2 \notag \\ 
	 & \quad + \tfrac{K}{2}\bigl( \nabla_{\beta}Q_{\alpha \beta}\bigr)^2
	   + \tfrac{K}{2} 
           \bigl( \epsilon_{\alpha \gamma \delta} \nabla_{\gamma} Q_{\delta \beta} 
           + 2q_0 Q_{\alpha \beta} \bigr)^2. 
\label{eq:FreeEnergy}
\end{align}
Here, $A_0$ sets the energy scale, $K$ is an elastic constant, $q_0=2\pi/p$ where $p$ is the cholesteric pitch, and $\gamma$ is a temperature-like control parameter governing proximity to the isotropic-cholesteric transition.
Greek indices denote Cartesian components and summation over repeated indices is implied; $\epsilon_{\alpha\gamma\delta}$ is the permutation tensor.
\if{
The velocity field obeys the continuity and Navier-Stokes equation, with a stress tensor generalised to describe liquid crystal hydrodynamics~\cite{beris1}. Within our hybrid scheme, we solve the Navier-Stokes equation via LB, and the Beris-Edwards equations for the evolution of $\mathbf{Q}$ (Eq. 1 in main text) via finite difference~\cite{juho1}.
Colloids are represented by the standard method of bounce-back on links, which leads to a no-slip boundary condition for the velocity field (see~\cite{juho1,ladd1} for details). Order parameter variations create an additional elastic force acting on the particle
which is computed by integrating the stress tensor over the particle surface~\cite{juho1,zumer_colloid}.
}\fi
The physics of our cholesteric dimers is determined by several dimensionless parameters. The chirality, 
\begin{equation}\label{eq:kappa}
  \kappa=\sqrt{\frac{{108 K q_0^2}}{{A_0 \gamma}}},
\end{equation}
and the reduced temperature, 
\begin{equation}\label{eq:tau}
\tau=\frac{27 (1-\gamma/3)}{\gamma}, 
\end{equation}
determine the equilibrium phase of the chiral nematic fluid~\cite{bp1}. 
We used the following parameters (in simulation units): $A_0=1.0,~K = W \simeq 0.065,~\xi=0.7,~\gamma = 3.0$, $p=32$, $q=1/2$ and $\Gamma =0.5$. 
Note that in our units the density is equal to unity, and so is the rotational
viscosity, $\gamma_1=\frac{2q^2}{\Gamma}$.
Our parameters yield $\tau=0$, $\kappa=0.3$. 
We used a periodic cubic simulation box
of volume $V=~192^3$ and particles of radius $R=7.25$. Note that the exact forms of Eqs.~S\ref{eq:kappa} and S\ref{eq:tau} follows the original work on Landau theory of blue phases by Grebel {\it et al.}~\cite{grebel}. The chirality, $\kappa$,  differs by a factor of $2$, from that used by Wright and Mermin~\cite{WrightMermin} as explained in the appendix A of~\cite{WrightMermin}. For more detailed treatment see eg.~\cite{gareth}.

We considered a degenarate planar anchoring of the director $\hat{\mathbf{n}}$ at the particle surface. Following Fournier and Galatola~\cite{galatola1} , we constructed an orientational order tensor, $\mathbf{Q}^{0}$, planar to the surface normal $\mathbf{\nu}$  as follows:
\begin{equation}
Q^{0}_{\alpha\beta}=\bigl(\delta_{\alpha\gamma}-\nu_{\alpha}\nu_{\gamma}\bigr)\tilde{Q}_{\gamma\delta}\bigl(\delta_{\delta\beta}-\nu_{\delta}\nu_{\beta}\bigr) - \tfrac{1}{3}q\delta_{\alpha\beta},
\end{equation}
where
\begin{equation}
\tilde{Q}_{\alpha\beta}= Q_{\alpha\beta} + \tfrac{1}{3}q\delta_{\alpha\beta}.
\end{equation}

\section{Calculation of the elastic pair potential}
The calculation of the elastic potential (Fig. 1a in the main text) was carried out by minimising the elastic free energy of equation~\ref{eq:FreeEnergy} by solving the Beris-Edwards equations~\cite{beris1} (Eq. (1) in the main text) in the absence of fluid flow. This was done for pairs of particles held static at separation $\mathbf{d}$. A first particle is
placed with its centre at the origin with the helical axis in the
$z$-direction. The far-field nematic director ($\hat{\mathbf n}(z)$) at
$z=0$ lies in the $x-$direction. A second particle is placed at
centre-centre separation vector $\mathbf{d}$ from the first. By repeating the simulations for different $\mathbf{d}$ an effective elastic two body potential can be mapped out.

We use a discrete representation of the colloidal particle surface in the lattice, which is a standard method in lattice Boltzmann models~\cite{juho1,ladd1}. This has a techical issue that the volume occupied of a particle can vary depending on where exactly the centre of the particle $\mathbf{r}$ is located with respect to the lattice. This can have a significant effect when considering the elastic interaction energy of colloidal pairs embedded in liquid crystal host.
To evaluate the error associated with the discretisation, we carried out a series of simulations where we generated the particle coordinates for the colloid pair as,
\begin{equation}
  \mathbf{r}_1 = \mathbf{r}_0 + \mathbf{u},
\end{equation}
and 
\begin{equation}
  \mathbf{r}_2 = \mathbf{r}_1 + \mathbf{d}.
\end{equation}
Here $\mathbf{r}_0$ denotes the origin and the components of $\mathbf{u}$ are uniformly distributed random numbers $u_{\alpha}\in[0,1]$, giving the position of the centre of the particle relative to the lattice. 

We carried out 8 independent scans consisting of 18 different separations along each of the three coordinate axis $x,~y,~z$.
Here it can be noted that for a constant $\mathbf{u}$, the total volume occupied by the particles ($V$) is constant for all the separations considered along the coordinate axes.

For the angular scan, we used 19 different angles, $\theta\in[0,180^\circ]$. Here, the situation is manifestly different: $V$ varies with angle and is symmetric around $\theta=90^\circ$ only for $u_{\alpha}=0$ or $1$. This gives a further uncertainty on the angular scan in the $x-y$ plane at $(d-2R)/p\approx 0.05$ shown in inset of Fig. 1a in the main text.
From physical arguments the elastic interaction energy in the chosen configuration, should be symmetric around $\theta=90^\circ$. We confirmed this by simulation with $u_{\alpha}=0$. However, for $u_{\alpha}\ne 0$, neither our individual angular scans nor the averaged curve were exactly symmetric. So, we further averaged the samples around $\theta=90^\circ$, using 16 independent samples for $\theta\in[0,90^\circ[$.

Fig. 1a in the main text shows the averaged interparticle elastic potential with descriptive error bars of 161$k_BT$, 161$k_BT$ and 156$k_BT$ for $x,~y$ and $z$, respectively, whereas for the averaged angular scan all error bars are shown. For both cases, these are estimated as the standard deviation of the mean.

\section{Dynamics of the dimer}
\begin{figure}
  \includegraphics[width=10cm]{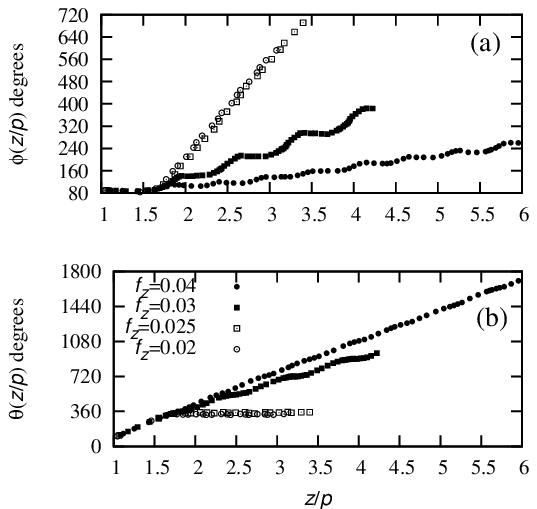}
  \caption{Replot of the data in Fig. 2e and 2f in the main text, against $z/p$. For the rotor motion (open symbols) only every 5th point is plotted for clarity.}
  \label{angles}
\end{figure}

To gain some deeper understanding of the dimer dynamics, we replot the time evolution of $\phi(t)$ and $\theta(t)$ (Fig. 2e and 2f in the main text) against the scaled distance along the helical axis $z/p$, shown in Fig. S\ref{angles}. From the $\phi(z/p)$ curve the rotor motion (Fig. S1a open symbols) is clear: after an initial transient, the colloid rotates following the constant twist of the cholesteric host.

\begin{figure}
  \includegraphics[width=8cm]{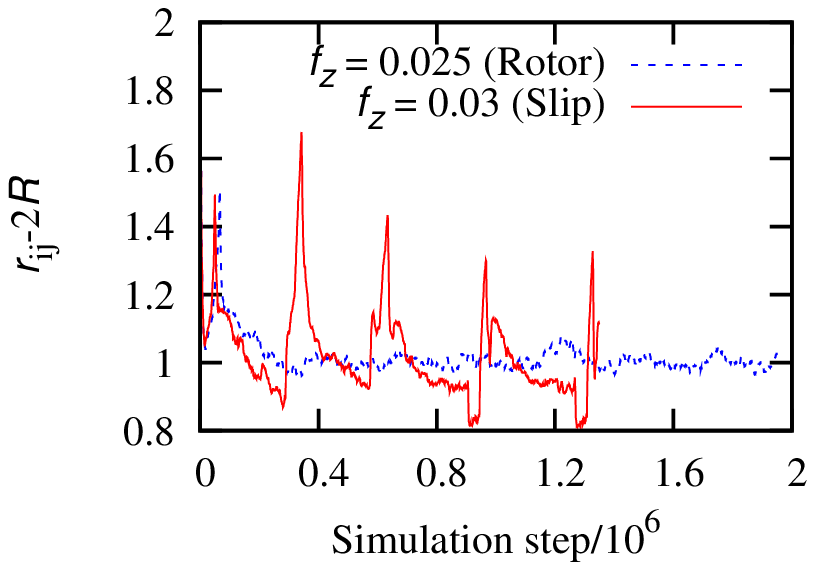}
  \caption{Time evolution of the surface-to-surface gap, $r_{ij} - 2R$, for the rotor (dashed line) and slip motion (solid line), respectively.}
  \label{gap}
\end{figure}

In the phase slippage regime, the colloidal dimer sediments along the helical axis with non-steady velocity: the motion along $z$ alternates between spells of slower and faster sedimentation. The motion is faster when the dimer sinks without rotating in the $x-y$ plane. This can be observed by comparing the time evolution of the angle $\phi(t)$ and phase lag $\theta(t)$ (solid symbols in Fig. 2e and 2f in the main text, respectively) with $\phi(z/p)$ and $\theta(z/p)$ (Fig. S1, solid symbols). In Fig. S1, intervals of constant $\phi$ denote sedimentation without rotation, whereas intervals of constant $\theta$ denote rotation locked to the cholesteric helix. By comparing $\phi(t)$ (Fig. 2e) with $\phi(z/p)$ (Fig. S1a), one may observe that intervals of rising $\phi$, corresponding to a roto-translation of the dimer, last longer in time than the constant $\phi$, but covers a similar distance in space -- therefore when rotating as a corkscrew the dimer takes longer to move the same distance, hence is slower.

During the sedimentation in the phase slippage regime, the dimer goes through a repulsive barrier in the elastic interaction potential at phase lag intervals of $\theta(t)=180^{\circ}$. This manifests itself in a repulsion between particles leading to a sharp increase of the surface-to-surface separation ($r_{ij}-2R$), as shown in Fig. S2 (solid line). For the rotor motion, after the initial transient period, the distance between the colloid pair stays constant (dashed line in Fig. S2).

\section{The rotor-phase slippage transition}
Finally, we comment in more detail on theories related to our dynamical transition between the rotor and the phase slip regimes. 
Firstly, our simplified Eqs. (3) may be seen as a relative of the Frenkel-Kontorova model, and of related models for the depinning of charge density waves in superconductors. The original Frenkel-Kontorova model considers the dynamics of a chain of forced coupled particles in a periodic potential (more realistic models for charge density wave depinning consider impurities hence an additional random potential). The key result of the simplest model is that there is a transition between pinned and moving particles. Just above the critical force, $f_c$, the asymptotic velocity of the particles, $v$, scales as~\cite{frenkel-kontorova1}:
\begin{equation}
v\sim \frac{\sqrt{f^2-f_c^2}}{\gamma}\propto\sqrt{f-f_c}
\end{equation}
where $f$ is the forcing, and $\gamma$ is an effective friction. The velocity therefore may be seen as an order parameter for the dynamical depinning transition, which is a continuous transition in this framework. While in Eqs. (3) in the main text there is a coupling between rotatory and translational motion, the analogue order parameter may be taken as the asymptotic time derivative of the lag angle between the dimer orientation and the local director field, $\theta(t)$, which we call $\Omega_{\rm lag}$. Via a numerical solution of Eqs. (3) we find that $\Omega_{\rm lag}\sim \sqrt{f^2-f_c^2}\propto \sqrt{f-f_c}$, for $f>f_c$, further validating the qualitative similarity hinted at above. In this analogy, our rotor phase corresponds to the pinned state, and the phase slip phase corresponds to the depinned one. The critical behaviour appears to be the same, provided the appropriate order parameters are chosen. 

\begin{figure}
  \includegraphics[width=8cm]{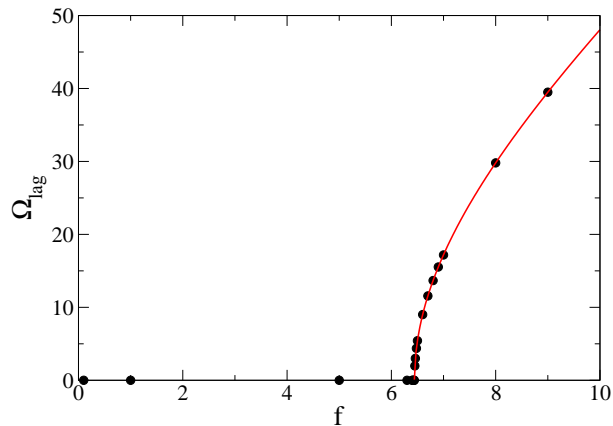}
  \caption{Plot of $\Omega_{\rm lag}$ as a function of $f$, for a numerical solution of Eqs. (3) in the main text ($\gamma_{\phi}=\gamma_z=1$, $q_0=2\pi$ in simulation units). The solid line is a fit to $\sqrt{f^2-f_c^2}$, leading to $f_c\sim 6.44$.}
  \label{gap2}
\end{figure}

Secondly, our model bears some similarities with the Kuramoto model~\cite{kuramoto1}, which is a paradigm to study the synchronisation of a system of oscillators. In the particularly simple case of two coupled oscillators, the equations describing the dynamics of the Kuramoto model are
\begin{eqnarray}
\frac{d\theta_1}{dt} & = & \omega_1+K/2\left[\sin\left(\theta_2-\theta_1\right)\right] \\
\frac{d\theta_2}{dt} & = & \omega_2+K/2\left[\sin\left(\theta_1-\theta_2\right)\right], 
\end{eqnarray}
where $\theta_1$ and $\theta_2$ describe the oscillators, $\omega_1$ and $\omega_2$ are the forcing for the two oscillators, and $K>0$ is the coupling constant favouring synchronisation ($\theta_1=\theta_2$). The relevant dynamical variable is $\psi=\theta_1-\theta_2$, which obeys the following evolution equation:
\begin{equation}
\frac{d\psi}{dt}  =  \Delta \omega-K\left[\sin\left(\psi\right)\right]. 
\end{equation}
where $\Delta\omega=\omega_1-\omega_2$. If we define an (asymptotic) angular velocity of the phase difference between the oscillators as $\Xi=\lim_{t\to \infty} \frac{d\psi}{dt}$, then $\Xi$ may be estimated as $2\pi$ divided by the time needed for $\psi$ to go from $\psi_0$ to $\psi_0+2\pi$ (this is independent of $\psi_0$), namely
\begin{eqnarray}
\Xi & = & \frac{2\pi}{\Delta t} \\
& = & \frac{2\pi}{\int_0^{2\pi} \frac{d\psi}{\Delta \omega-K\sin(\psi)}} \\
& = & \sqrt{\left(\Delta\omega^2-K^2\right)}
\end{eqnarray}
where the final integral can be done with the method of residues, by changing
variables to the complex number $z=e^{i\psi}$ and noting that 
$\sin(\psi)=\frac{e^{i\psi}-e^{-i\psi}}{2i}$. As in our simplified model, 
we find that the relevant ``order parameter'', $\Xi$, behaves as $\sqrt{\Delta\omega^2-\Delta\omega_c^2}$, with $\Delta\omega_c=K$, close to the transition.

Finally, the full lattice Boltzmann simulations of our cholesteric dimers lead to a similar scaling for $\Omega_{\rm lag}$ (see Fig. 3c in the main text). The critical force estimated through the fit is $f_c\approx 0.0267\pm 0.0005$, which is in agreement with $f_c\in]0.025,0.0275[$ observed in simulations. This shows that our minimal theory captures the physics of the dynamical phase transition between the rotor and slip phases, and also suggests that this is in the same universality class of oscillator synchronisation and of the depinning of charge density waves in a perfectly periodic medium. 

Note that, although the near-critical behaviour of our ``order parameter'' is given by $\sim\sqrt{f-f_c}$, both our simulation data and the numerics corresponding to Eqs. (3) in the main text are better fitted by a function proportional to $\sqrt{f^2-f_c^2}$, as this better captures the behaviour of $\Omega_{\rm lag}$ further away from $f_c$.


\begin{thebibliography}{99}
\bibitem{colloidal_walkers} C. E. Singa {\it et al.}, 
{\it Proc. Natl. Acad. Sci. USA} {\bf 107}, 535 (2010);
P. Tierno {\it et al.}, 
{\it Phys. Rev. Lett.} {\bf 101}, 218304 (2008).
\bibitem{nematic_self_assembly} M.~Ravnik {\it et al.}, {\it Phys. Rev. Lett.}
{\bf 99}, 247801 (2007); J.~C.~Loudet, {\em et al.}, {\it Langmuir} {\bf 20}, 
11336 (2004).
\bibitem{Tanaka}
T.~Araki, H.~Tanaka, {\it Phys. Rev. Lett.} {\bf 97}, 127801 (2006).
\bibitem{zumer_colloid}M.~Skarabot {\em et al.} 
{\it Phys. Rev. E} {\bf 77}, 061706 (2008);  M.~Conradi {\em et al.},  
{\it Soft Matter} {\bf 5}, 3905 (2009).
\bibitem{juho} J. S. Lintuvuori {\it et al.}, {\it Phys. Rev. Lett.} {\bf 105}, 178302 (2010); {\it J. Mat. Chem.} {\bf 20}, 10547 (2010).
\bibitem{colin} F.~E. Mackay and C.~Denniston, {\it EPL} {\bf 94}, 66003 (2011).
\bibitem{lavrentovich_planar} I.~I.~Smalyukh {\it et al.}, {\it Phys. Rev. 
Lett.} {\bf 95}, 157801 (2005).
\bibitem{frenkel-kontorova}
J. Frenkel and T. Kontorova, {\it Zh. Eksp. Teor. Fiz.} {\bf 8}, 1340
(1938); R. Besseling. R. Niggebrugge and P. H. Kes, {\it Phys. Rev. Lett.} {\bf 82}, 3144 (1999); R. Besseling {\em et al.}, {\it Europhys. Lett.} {\bf 62}, 419 (2003).
\bibitem{kuramoto} J. A. Acebron {\em et al.}, 
{\it Rev. Mod. Phys.} {\bf 77}, 137 (2005).
\bibitem{beris} 
A.~N.~Beris and B.~J.~Edwards, {\it Thermodynamics of Flowing Systems},
Oxford University Press, Oxford, (1994).
\bibitem{supmat} See online Supplementary Material at XXX for additional details on our model and dynamical transition and for movies of the dynamics.
\bibitem{galatola} J.-B.~Fournier, P.~Galatola, {\it Europhys. Lett.} {\bf 72}, 403 (2005).
\bibitem{LBLC}
D.~Marenduzzo et al., {\it Phys. Rev. E} {\bf 76}, 031921 (2007); M.~E.~Cates et al., {\it Soft Matter} {\bf 5}, 3791 (2009).
\bibitem{ladd} N.-Q. Nguyen and A. J. C. Ladd, {\it Phys. Rev. E}
{\bf 66}, 046708 (2002).
\bibitem{mozaffari} M.~R.~Mozaffari {\it et al.}, {\it Soft Matter} {\bf 7}, 1107 (2011).
\bibitem{twisted_nem} U.~Tkalec {\it et al.}, {\it Phys. Rev. Lett.} {\bf 103}, 127801 (2009).
\bibitem{note}$\theta \approx 348^{\circ}$ and $\theta \approx 330^{\circ}$, for $f=0.25$ and $f=0.2$, respectively. These correspond to $\theta = 12^{\circ}$ and $\theta = 30^{\circ}$ in the inset of Fig.~1a. 
\bibitem{2sphere_sedimentation} M. Stimpson, G. B. Jeffery, 
{\it Proc. Roy. Soc. A} {\bf 111}, 110 (1926). 
\bibitem{two-point-microrheology} A.~S.~Khair and T.~M.~Squires, {\it Phys. Rev. Lett.} {\bf 105}, 156001 (2010).
\bibitem{platelets_electric} C. P. Lapointe {\it et al.}, {\it Phys. Rev. Lett.} {\bf 105}, 178301 (2010). 
\bibitem{lavrentovich_backflow} O. P. Pishnyak {\em et al.}, 
{\it Phys. Rev. Lett.} {\bf 106}, 047801 (2011);
O. P. Lavrentovich {\em et al.}, {\it Nature} {\bf 467}, 947 (2011).
\end{thebibliography}

\begin{thebibliography}{99}

\bibitem{beris1} 
A.~N.~Beris and B.~J.~Edwards, {\it Thermodynamics of Flowing Systems},
Oxford University Press, Oxford, (1994).
\bibitem{bp1} O. Henrich {\it et al.}, {\it Phys. Rev. E} {\bf 81}, 031706 (2010); {\it Proc. Natl. Acad. Sci. USA} {\bf 107}, 13212 (2010).
\bibitem{grebel}
H.~Grebel, R.~M.~Hornreich and S.~Shtrikman, {\it Phys. Rev. A} {\bf 28}, 114 (193).
\bibitem{WrightMermin}
D.~C.~Wright and N.~D.~Mermin, {\it Rev. Mod. Phys.} {\bf 61}, 385 (1989).
\bibitem{gareth}
G.~P.~Alexander and J.~M.~Yeomans, {\it Phys. Rev. E} {\bf 74}, 061706 (2006).
\bibitem{galatola1} J.-B.~Fournier, P.~Galatola, {\it Europhys. Lett.} {\bf 72}, 403 (2005).
\bibitem{juho1} J. S. Lintuvuori {\it et al.}, {\it Phys. Rev. Lett.} {\bf 105}, 178302 (2010); {\it J. Mat. Chem.} {\bf 20}, 10547 (2010).
\bibitem{ladd1} N.-Q. Nguyen and A. J. C. Ladd, {\it Phys. Rev. E}
{\bf 66}, 046708 (2002).
\bibitem{frenkel-kontorova1}
 J. Frenkel and T. Kontorova, {\it Zh. Eksp. Teor. Fiz.} {\bf 8}, 1340
(1938); R. Besseling. R. Niggebrugge and P. H. Kes, {\it Phys. Rev. Lett.} {\bf 82}, 3144 (1999); R. Besseling {\em et al.}, {\it Europhys. Lett.} {\bf 62}, 419 (2003).
\bibitem{kuramoto1} J. A. Acebron, L. L. Bonilla, C. J. Perez Vicente, F. Ritort and R. Spigler, {\it Rev. Mod. Phys.} {\bf 77}, 137 (2005). 
\end{thebibliography}
\end{document}